\documentclass[sigconf]{acmart}

\usepackage{multirow}
\usepackage{subcaption}
\usepackage{booktabs}
\usepackage{enumerate}
\usepackage{makecell}
\usepackage{tikz}
\usepackage{enumitem}
\usepackage{graphicx}
\usepackage{bm}
\usepackage[table]{xcolor}
\usepackage{algorithm}
\usepackage{algpseudocode}

\newcommand{\LongState}[1]{%
  \State \parbox[t]{\dimexpr\linewidth-\algorithmicindent\relax}{#1}%
}
\AtBeginDocument{%
  }

\copyrightyear{2026}
\acmYear{2026}
\setcopyright{cc}
\setcctype{by}
\acmConference[SIGIR '26] {Proceedings of the 49th International ACM SIGIR Conference on Research and Development in Information Retrieval}{July 20--24, 2026}{Melbourne, VIC, Australia.}
\acmBooktitle{Proceedings of the 49th International ACM SIGIR Conference on Research and Development in Information Retrieval (SIGIR '26), July 20--24, 2026, Melbourne, VIC, Australia}
\acmISBN{979-8-4007-2599-9/2026/07}
\acmDOI{10.1145/XXXXXX.XXXXXX}

\begin{document}

\title{DIAURec: Dual-Intent Space Representation Optimization for Recommendation}


\author{Yu Zhang}
\affiliation{%
  \institution{Anhui University}
  \city{Hefei}
  \country{China}
}
\email{zhangyu.ahu@gmail.com}

\author{Yiwen Zhang}
\authornote{Yiwen Zhang is the corresponding author.}
\affiliation{%
  \institution{Anhui University}
  \city{Hefei}
  \country{China}
}
\email{zhangyiwen@ahu.edu.cn}

\author{Yi Zhang}
\affiliation{%
  \institution{Anhui University}
  \city{Hefei}
  \country{China}
}
\email{zhangyi.ahu@gmail.com}

\author{Lei Sang}
\affiliation{%
  \institution{Anhui University}
  \city{Hefei}
  \country{China}
}
\email{sanglei@ahu.edu.cn}


\renewcommand{\shortauthors}{Yu Zhang, Yiwen Zhang, Yi Zhang, and Lei Sang.}

\begin{abstract}
General recommender systems deliver personalized services by learning user and item representations, with the central challenge being how to capture latent user preferences. 
However, representations derived from sparse interactions often fail to comprehensively characterize user behaviors, thereby limiting recommendation effectiveness. 
Recent studies attempt to enhance user representations through sophisticated modeling strategies ($e.g.,$ intent or language modeling). Nevertheless, most works primarily concentrate on model interpretability instead of representation optimization. 
This imbalance has led to limited progress, as representation optimization is crucial for recommendation quality by promoting the affinity between users and their interacted items in the feature space, yet remains largely overlooked. 
To overcome these limitations, we propose DIAURec, a novel representation learning framework that unifies intent and language modeling for recommendation.  
DIAURec reconstructs representations based on the prototype and distribution intent spaces formed by collaborative and language signals. 
Furthermore, we design a comprehensive representation optimization strategy. 
Specifically, we adopts \textit{alignment} and \textit{uniformity}  as the primary optimization objectives, and incorporates both coarse- and fine-grained matching to achieve effective alignment across different spaces, thereby enhancing representational consistency. 
Additionally, we further introduce intra-space and interaction regularization to enhance model robustness and prevent representation collapse in reconstructed space representation. 
Experiments on three public datasets against fifteen baseline methods show that DIAURec consistently outperforms state-of-the-art baselines, fully validating its effectiveness and superiority. 
\end{abstract}




\maketitle

\section{Introduction} 
With the explosive growth of web data, diverse Internet services, including video and music streaming, have become deeply integrated into people’s daily lives \cite{peroni_2025_video}. 
As a core component of online platforms, recommender systems \cite{Wu_SurveyRecSys_TKED_2023,gao_gnn2-survey_RS_2023} are essential for delivering personalized experiences. 
Traditionally, recommendation models have primarily relied on historical interaction data, using diverse neural networks architectures \cite{fan_www_nnRS_2019,he_WWW_NCF_2017,wang_SIGIR_NGCF_2019} to learn collaborative representations. 
However, in practice, user preferences are inherently complex and multifaceted, making it difficult for a single modeling paradigm to capture them adequately and thus limiting final recommendation quality. 

\begin{figure*}[t]
\centering
\includegraphics[width=1.0\textwidth]{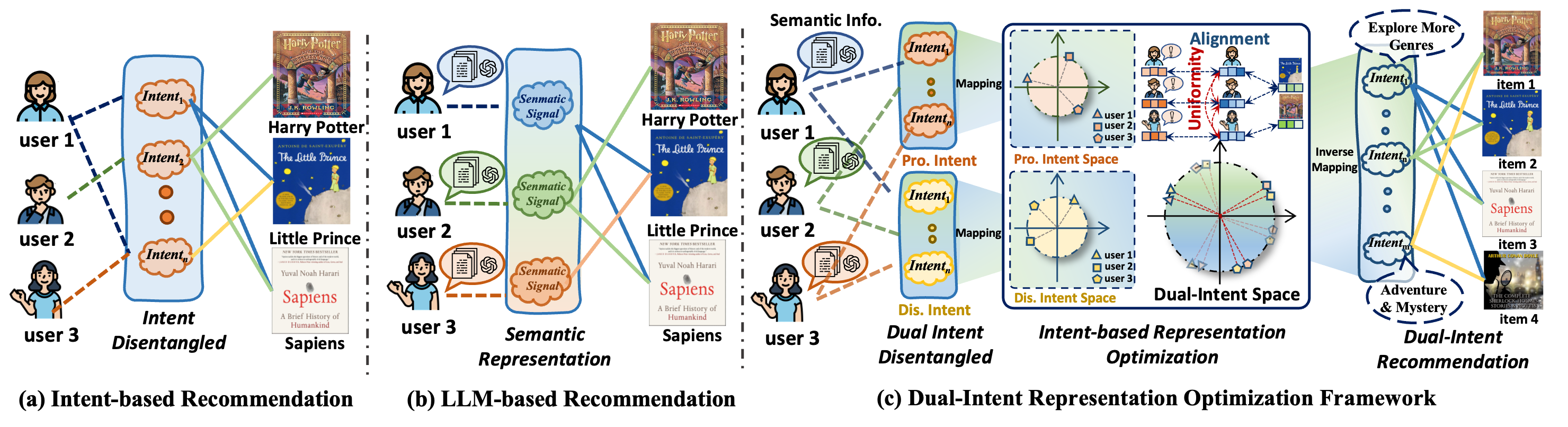} 
\caption{(a) Intent-based recommendation paradigm. 
(b) LLM-based recommendation paradigm. (c) Representation reconstruction through dual-intent space (\textit{i.e.,} prototype and distribution intents), with a comprehensive optimization strategy for personalized recommendation.}
\label{fig:motivation}
\end{figure*}

Amid this trend, a series of modeling methods \cite{liu_KDD_modelsurvey_2024,ko_Electronics_modelsurvey_2022} have emerged. 
Several pioneering studies introduce the notion of \textit{intent} \cite{wang_SIGIR_DGCF_2020,DCCF_2023_SIGIR} to disentangle users’ potential behaviors. 
As shown in Fig. \ref{fig:motivation}(a), intent-based methods differ from conventional paradigms by projecting user–item interactions into an intent space. 
These intents are subsequently integrated with collaborative representations to provide more personalized recommendations. 
Representative works such as DCCF \cite{DCCF_2023_SIGIR} and BIGCF \cite{zhang_BIGCF_SIGIR_2024} perform fine-grained disentanglement of multiple user and item intents, thereby improving both the accuracy and interpretability of recommendation results. Meanwhile, the rapid progress of large language models (LLMs) \cite{wu_LLMsurvey_2024,wang2026mllmrec,wang2025multimodal} have enabled the construction of rich contextual semantic representations. 
As illustrated in Fig.~\ref{fig:motivation}(b), LLM-based methods \cite{RLMRec_2024_WWW,KAR_2024_RS} often leverage an LLM to construct user and item profiles, and model semantic representations from contextual information, thereby providing supplementary semantic signals for recommendation models. 

Despite these advances, most research still suffers from the following limitations. \textbf{On the one hand}, most intent-based methods \cite{wang_SIGIR_DGCF_2020,DCCF_2023_SIGIR} are fundamentally rely on interaction data. 
Specifically, these methods capture latent information by modeling intent space to construct more interpretable representations. 
However, to the best of our knowledge, the disentanglement of multiple intents is almost entirely based solely on observed interactions \cite{DCCF_2023_SIGIR,wang_SIGIR_DGCF_2020}. 
In scenarios with sparse user behavior, relying on interactions makes it difficult to obtain robust representations \cite{chen_2025_datasparsity}. 
Furthermore, even when sufficient interactions are available, mainstream methods often model intent representation as deterministic point vectors, neglecting the inherent noise and uncertainty present in interaction data \cite{jiang_uncertainty_ijcai_2019}. 
Therefore, such intent representations may fail to capture the true underlying user preferences, leading to representation drift and instability. 

\textbf{On the other hand}, LLM-based methods substantially increase the computational burden \cite{sui_2025_llcomputational}. Due to LLM's considerable memory and processing requirements, the integration of LLM into recommender systems inevitably amplifies training and inference costs. 
Recent studies \cite{zhao_TKDE_llmsurvey_2024,RLMRec_2024_WWW} employ LLM to model semantic representation that enrich overall representation. 
However, modeling semantic representations is becoming increasingly computationally expensive, particularly when relying on complex prompt designs \cite{2025IRLLRec} or chain-of-thought strategies \cite{yue_AAAI_cot4rec_2025}, thereby incurring substantial computational overhead. 

\textbf{Finally}, and most critically, we argue that representation optimization has not received sufficient attention. 
Existing researchers \cite{chen_WWW_IntentModeling_2022,RLMRec_2024_WWW,jiang_KDD_adaptive_2023} have explored interpretable modeling strategies from multiple perspectives to enhance collaborative representations. Undeniably, these methods have improved recommendation performance to some extent, primarily due to the additional supervisory signals. 
However, the key question is whether these representations have truly realized their full potential. 
We contend that differences in modeling methods often lead to inconsistencies among the constructed representations. If these representations are directly introduced without considering their joint optimization, they may suffer from representation collapse \cite{chen_WSDM_collapse_2024,peng_TOIS_collapse_2025}. 
Over the long term, this neglect may weaken the essential role of representation optimization in recommendation and constrain further improvements in overall performance. 
Therefore, this work aims to explore whether it is possible to achieve the following objectives simultaneously: 

\begin{itemize}[leftmargin=*,label=--]
    \item How can intent representations be modeled more robustly while alleviating interaction sparsity and uncertainty? 
    \item How can semantic signals be effectively leveraged to enrich representations without incurring substantial computational cost? 
    \item How can collaborative optimization across different representation spaces be realized to enhance recommendation quality?
\end{itemize}

To achieve the above objectives, we propose a novel \textbf{D}ual \textbf{I}ntent representation learning framework with \textbf{A}lignment and \textbf{U}niformity for \textbf{Rec}ommendation (\textbf{DIAURec}). 
Considering the sparsity of user interactions, we incorporate LLM to enrich semantic signals and model them as prototype intent, while we introduce distribution intent on the collaborative side to capture latent representation distribution. 
This dual intent modeling supplies more informative signals for representation learning. 
Nevertheless, as noted earlier, the central limitation concerns achieving effective coordination across different representation spaces. Motivated by this limitation, we design a comprehensive representation optimization strategy that adopts \textit{alignment} and \textit{uniformity} \cite{wang_ICML_understand_2020} as the primary objectives. 
However, relying solely on this paradigm remains insufficient for high-quality collaborative optimization. 
Building upon this, we further refine the uniformity objective and introduce both coarse-grained and fine-grained matching approaches, coupled with two regularization technologies, to enable effective cross-space alignment and integration while mitigating representation collapse. As illustrated in Fig. \ref{fig:motivation}(c), the overall optimization paradigm of DIAURec is conceptually illustrated. 

The contributions of this work can be summarized as follows:

\begin{itemize}[leftmargin=*,label=--]
    \item We propose a novel recommendation framework called DIAURec, which reconstructs collaborative representations by modeling prototype intent and distribution intent, thereby capturing both uncertain and semantic preference signals.

    \item We design a integrated representation optimization strategy, which adopts \textit{alignment} and \textit{uniformity} as the primary objectives, and further introduces coarse-grained and fine-grained  matching approaches with two regularization techniques to achieve joint optimization across different spaces.

    \item We conduct extensive experiments on three real-world datasets against fifteen baseline methods from three categories, and the results consistently demonstrate the superiority and effectiveness of DIAURec.
\end{itemize}

\section{Methodology}

\begin{figure*}[t]
\centering
\includegraphics[width=1.0\textwidth]{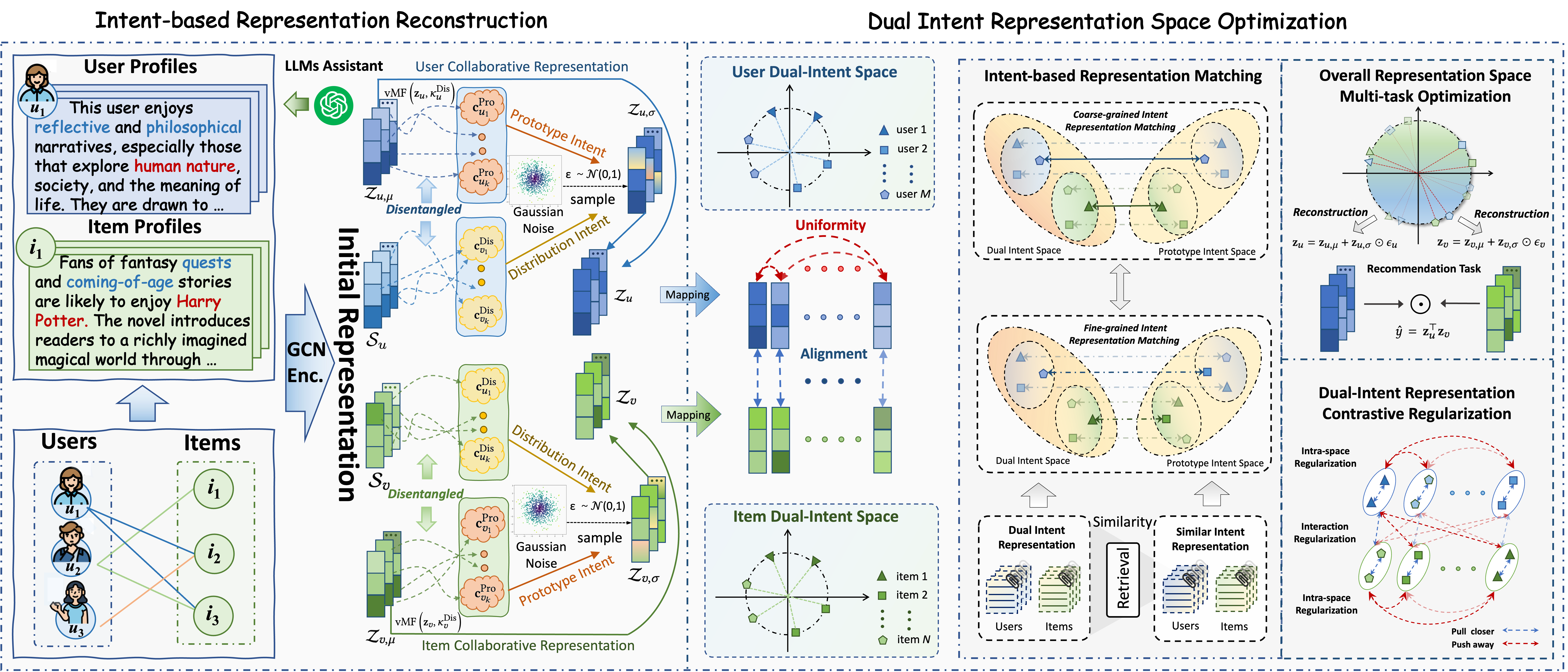} 
\caption{The complete framework of the proposed DIAURec. DIAURec contains two major components: (i) Prototype- and distribution-intent reconstruction for representations, and (ii) Comprehensive optimization of representation spaces with alignment–uniformity objectives, fine- and coarse-grained matching, and regularization technologies for recommendation. }
\label{fig:framework}
\end{figure*}

\subsection{Problem Definition}

Without loss of generality, a recommendation task involves a set of $M$ users $\mathcal{U}=\{u_1,\dots,u_M\}$ and a set of $N$ items $\mathcal{V}=\{v_1,\dots,v_N\}$ \cite{he_WWW_NCF_2017}. 
Historical interactions are represented by a binary matrix $\mathbf{R}\in \{0,1\}^{M\times N}$, where $R_{uv}$=1 if user $u$ has interacted with item $v$, and $R_{uv}$=0 otherwise \cite{wang_SIGIR_NGCF_2019,he_LightGCN_SIGIR_2020}. 
The goal of recommendation is to learn a function $f(\mathcal{U}\times\mathcal{V})\to \mathbf{\hat{R}}$ that predicts the preference score $\hat{y}_{uv}$ for unobserved user–item pairs. 
To capture high-order structural signals, we adopt graph convolutional (GCN) paradigm \cite{sharma_CS_graphsurvey_2024} in which representations are iteratively propagated: $\mathbf{Z}^{(l+1)} = \mathbf{D}^{-\frac{1}{2}} \mathbf{A} \mathbf{D}^{-\frac{1}{2}} \mathbf{Z}^{(l)},$ 
where $l$ is the GCN layer, $\mathbf{A}$ is the adjacency matrix and $\mathbf{D}$ is the corresponding degree matrix. After $L$ layers, we obtain the user and item representations $\mathbf{z}_u, \mathbf{z}_v \in \mathbb{R}^{1\times d}$, respectively. The predicted preference score is then given by the inner product:
$\hat{y} = \mathbf{z}_u^\top \mathbf{z}_v$. 
Following this formulation, we present DIAURec, the proposed dual-intent representation learning framework, as shown in Fig. \ref{fig:framework}. 

\subsection{Representation Reconstruction Module}

Before performing recommendation, obtaining high-quality representations is equally crucial. 
Therefore, we revisit representation modeling from the hyperspherical perspective. 
Specifically, each observed interaction $(u,v)$ is regarded as a Bernoulli trial \cite{mnih_PMF_2007}, whose probability is determined by user and item representations constrained on the hypersphere \cite{wang_ICML_understand_2020}: $\mathbb{P}(\hat R_{uv}=1 \mid \mathbf z_u, \mathbf z_v) 
= \sigma\!\left(\mathbf z_u^\top \mathbf z_v\right). $ 
Accordingly, overall likelihood of observed interactions is given by: 
\begin{equation}
\mathcal L(\hat {\mathbf R};\, \mathcal{Z}, \Theta)
= \prod_{(u,v)\in\mathcal E}
\mathbb{P}\!\left( \hat R_{uv}\,\middle|\,\mathbf z_u,\mathbf z_v;\,\Theta\right),
\end{equation}
where $\mathcal E$ denotes the set of observed user–item interaction pairs. Subsequently, the key challenge shifts to the modeling and reconstruction probabilistic representations, so that we can effectively capture both collaborative and semantic signals.

\subsubsection{ \textbf{Representation Modeling.}} 
We first briefly present the construction of collaborative and semantic representations. 

\noindent\textbf{Collaborative Modeling.} 
Similar to previous studies \cite{wu_SIGIR_SGL_2021,Yu_Unveiling_2025}, each user and item is associated with ID information. 
Here, we model collaborative representations in a probabilistic manner, where each representation is expressed as a Gaussian distribution \cite{jiang_uncertainty_ijcai_2019,liang_WWW_VAE_2018}: 
\begin{equation}
\label{eq:col_model}
\mathbf{z}_u \sim \mathcal{N}(\boldsymbol\mu_u, \mathrm{diag}[\boldsymbol\sigma_u^2]), 
\mathbf{z}_v \sim  \mathcal{N}(\boldsymbol\mu_v, \mathrm{diag}[\boldsymbol\sigma_v^2]),
\end{equation}
where $\boldsymbol\mu_u$ and $\boldsymbol\mu_v$ are the approximate mean, and $\boldsymbol\sigma_u^2$ and $\boldsymbol\sigma_v^2$ denote the approximate variances. 
This distributional formulation provides more flexible space for capturing collaborative information. 

\noindent\textbf{Semantic Modeling.} 
Beyond collaborative signals, users and items also contain rich context information, such as profiles or textual descriptions \cite{Zhang_WSDM_context_2017}. Leveraging large language models (LLM) \cite{wu_LLMsurvey_2024}, we aggregate multiple types of textual information $T_k(\cdot)$ and feed them into LLM to obtain unified profiles $\mathcal{P}(u;v) = \mathcal{F}_{\text{LLM}}(T_k(u;v)),$ where $\mathcal{F}_{\text{LLM}}(\cdot)$ denotes the profile generator from LLM. 
These profiles are then decoded 
$ \mathbf{\hat s}_u = f_{\text{LLM}}(\mathcal{P}(u)), 
\mathbf{\hat s}_v = f_{\text{LLM}}(\mathcal{P}(v)).$ 
Here, $f_{\text{LLM}}(\cdot)$ transforms profiles into dense semantic representations \cite{Bert,neelakantan_text_2022}. Followed by projection into the representation space as follows: 
\begin{equation}
\label{eq:sem_model}
\mathbf s_u = f_{\text{sem}}(\mathbf{\hat{s}}_u), 
\mathbf s_v = f_{\text{sem}}(\mathbf{\hat{s}}_v),
\end{equation}
where $f_{\text{sem}}(\cdot)$ denotes a multi-layer perception that reduces the dimensionality to match the collaborative space.

\subsubsection{\textbf{Intent Modeling.}} 

In recommender systems, users pursue diverse intents that drive their behaviors, even under similar interaction contexts. 
To this end, we explore different representations to disentangle distinctive intents. 

\noindent\textbf{Prototype Intent.} 
Semantic representations encode rich preferences and attributes. Therefore, we model prototype intents by associating semantic representations with a set of intent prototypes. Formally, the probability of assigning a user or item to the $k$-th prototype is defined as follows:

\begin{equation}
\label{eq:pro_intent_model}
\mathbb{P}(\mathbf{c}^\text{Pro}_k | \mathbf{s}_u) = \frac{\varphi(\mathbf{s}_u^\top \mathbf{c}^\text{Pro}_k / \eta )}{\sum^{\mathcal{K}}_{k'} \varphi(\mathbf{s}_u^\top \mathbf{c}^\text{Pro}_{k'}/ \eta)}, 
\mathbb{P}(\mathbf{c}^\text{Pro}_k | \mathbf{s}_v) = \frac{\varphi(\mathbf{s}_v^\top \mathbf{c}^\text{Pro}_k / \eta)}{\sum^{\mathcal{K}}_{k'} \varphi(\mathbf{s}_v^\top \mathbf{c}^\text{Pro}_{k'} / \eta)},
\end{equation}
where  $\varphi(\cdot)=\exp(\cdot)$, $\mathbf{c}^\text{Pro}_k$ denotes the $k$-th intent prototype, $\eta$ is the intent contribution coefficient, and $\mathcal{K}$ is the prototype set.

\noindent\textbf{Distribution Intent.} Unlike prototype intent, collaborative representations are learned directly from interaction data, which inevitably carry uncertainty and noise. To better capture potential preferences, we map collaborative representations onto the hypersphere via the von Mises–Fisher (vMF) distribution \cite{davidson_vMF_2018,Banerjee_vMF_2005}: 
\begin{equation}
\label{eq:map_distribution}
\mathbf{h}^{\text{Dis}}_u \sim \text{vMF}\left( \mathbf{z}_u, \kappa^{\text{Dis}}_u \right), 
\mathbf{h}^{\text{Dis}}_v \sim \text{vMF}\left( \mathbf{z}_v, \kappa^{\text{Dis}}_v \right),
\end{equation}
where $\kappa^{\text{Dis}}$ denotes the concentration parameter. 
Subsequently, to align sampled directions with distribution intent, we compute an affinity score $\Psi_u = \mathbf{h}^{\text{Dis}}_u \cdot \mathbf{W}^{\text{Dis}},$ where  $\mathbf W^{\text{Dis}}\in\mathbb R^{\mathcal{K} \times d}$ is a projection matrix. 
Given this score, the probability of assigning user/item to the $k$-th distribution intent is defined as: 
\begin{equation}
\label{eq:dis_intent_model}
\mathbb{P}(\mathbf{c}^\text{Dis}_k | \mathbf{z}_u) = \frac{\varphi(\Psi_k / \eta)}{\sum^{\mathcal{K}}_{k'} \varphi(\Psi_{k'} / \eta)}, 
\mathbb{P}(\mathbf{c}^\text{Dis}_k | \mathbf{z}_v) = \frac{\varphi(\Psi_k / \eta)}{\sum^{\mathcal{K}}_{k'} \varphi(\Psi_{k'} / \eta)}.
\end{equation}

\subsubsection{\textbf{Representation Reconstruction.}}
Building on above modeling, we aim to reconstruct the user and item representations \cite{liang_WWW_VAE_2018}. Specifically, we treat the collaborative representations as probability distributions $\mathbf{z}_{\mu}$, while the mixture of prototype and distribution intents is represented by $\mathbf{z}_{\sigma}$. 
We assume that each user–item interaction is jointly driven by prototype and distribution intents. The reconstructed representations are defined as follows: 
\begin{equation}
\label{eq:reconstruction_gcn}
\mathbf{z}_{x} = \frac{1}{L}\sum_{l \in L} \frac{1}{|\mathcal N_u||\mathcal N_v|} \mathbf{z}^{(l)}_{x,\mu} + (\mathbf{z}^\text{Pro}_{x,\sigma} + \mathbf{z}^\text{Dis}_{x,\sigma}) \odot \epsilon_{x}, 
\end{equation}
where $L$ is the GCN layers, the users and items is $x \in \{u,v\}$, $|\mathcal N_u|$ and $|\mathcal N_v|$ are the product of the cardinalities of the user and item neighborhoods, and $\mathbf{z}^\text{Pro}_{x,\sigma}$ and $\mathbf{z}^\text{Dis}_{x,\sigma}$ denote as follows: 
\begin{equation}
\label{eq:both_intent}
\mathbf{z}^\text{Pro/Dis}_{x,\sigma}=\mathbf{c}^{\text{Pro/Dis}}_{x} = \sum_{k \in \mathcal{K}} \mathbf{c}^\text{Pro/Dis}_k \cdot \mathbb{P} (\mathbf{c}^\text{Pro/Dis}_k \mid \mathbf{s}_{x};\mathbf{z}_x),    
\end{equation}

where $\mathbf{c}^\text{Pro}_k$ and $\mathbf{c}^\text{Dis}_k$ denote the $k$-th prototype and distribution intents. 
Accordingly, the interaction probability is obtained by marginalizing over reconstructed collaborative representation:
\begin{equation}
\label{eq:reconstruction_overall}
\mathbb{P}(\hat R_{uv}\mid \mathbf z_x,\mathbf s_x) 
= \sum_{k \in \mathcal{K}} \mathbb{P}(\hat R_{uv}\mid \mathbf z_x,\mathbf c_k)\,
\mathbb{P}(\mathbf{c}^\text{Pro}_k\mid \mathbf z_x)\,
\mathbb{P}(\mathbf{c}^\text{Dis}_k\mid \mathbf s_x).
\end{equation}
Here, $\mathbb{P}(\hat R_{uv}\mid \mathbf z_x,\mathbf c_k)$ denotes the intent-specific likelihood of the interaction, while $\mathbb{P}(\mathbf{c}^\text{Pro}_k\mid \mathbf z_x)$ and $\mathbb{P}(\mathbf{c}^\text{Dis}_k\mid \mathbf s_x)$ serve as prototype and distribution intent priors inferred from collaborative and semantic views. 
By mixing the dual intents, the interaction probability is reconstructed as a weighted combination of multiple factors, which disentangles diverse user behaviors and enhances expressiveness. 
The overall likelihood is finally expressed as follows: 
\begin{align}
\mathcal L(\hat{\mathbf R}; \mathcal Z, \mathcal S, \Theta)
& \sim 
\prod_{(u,v)\in \mathcal E} 
\mathbb{P}(\hat R_{uv}\mid \mathbf z_{x, \mu},\mathbf z_{x,\sigma}; \Theta) 
\nonumber \\
& \approx
\prod_{(u,v)\in \mathcal E} 
\mathbb{P}(\hat R_{uv}\mid \mathbf{z}_x, \mathbf c^\text{Pro}_x,\mathbf c^\text{Dis}_x; \Theta).
\end{align}

\subsection{Representation Optimization Module}
In this section, we propose alignment and uniformity as main optimization objectives. Additionally, we introduce diverse matching approaches and regularization techniques to promote representation quality and prevent representation collapse. 

\subsubsection{\textbf{Alignment and Uniformity}}
Recent studies \cite{wang_ICML_understand_2020,gui_TPAMI_survey_2024} demonstrate that two desirable properties—\textit{alignment} and \textit{uniformity}—serve as principled criteria for characterizing the distribution of representations on the hypersphere, providing theoretical guidance for representation optimization in downstream tasks. In recommendation scenarios, existing methods \cite{DirectAU_wang_2022_KDD,SIURec} explicitly treats these two properties as training objectives. The alignment and uniformity losses are defined as follows: 
\begin{equation} 
\label{eq:align}
\mathcal{L}_{\text{Align}}(\mathbf{e}_u,\mathbf{e}_v) = \mathbb{E}_{\langle u,v \rangle \in \mathcal{B}_{p}} \|\mathbf{e}_u-\mathbf{e}_v\|_2^2,
\end{equation}
\begin{align}
\label{eq:uniform}
\mathcal{L}_{\text{Uniform}}(\mathbf{e}_u;\mathbf{e}_v) = 
&\log \mathbb{E}_{u,u' \sim \mathcal{B}_{u}}  \varphi (-2\|\mathbf{e}_{u}-\mathbf{e}_u'\|_2^2) \ \ + 
\nonumber \\ 
&\log \mathbb{E}_{v,v' \sim \mathcal{B}_{v}} \varphi (-2\|\mathbf{e}_{v}-\mathbf{e}_v'\|_2^2), 
\end{align}
where $\varphi(\cdot)=\exp(\cdot)$, $\mathcal{B}_p$ is the set of positive pairs in the batch, and $\mathcal{B}_u, \mathcal{B}_v$ are the user and item samples. Intuitively, alignment encourages positive pairs to be pulled closer, while uniformity ensures that user and item representations are uniformly distributed.

Despite the above losses are effective for representation optimization, joint optimization across distinct spaces remains significant challenge. 
Specifically, independent optimization within each space often leads to difficulties in matching as well as distribution shifts, thereby weakening the effectiveness of model optimization. 
To this end, we propose the reconstructed representations $\mathbf{z}_u, \mathbf{z}_v$ for direct optimization, with the primary objective defined as follows:

\begin{equation}
\label{eq:item_uniform}
    \mathcal{L}_{\text{AU}} = \mathcal{L}_\text{Align}(\mathbf{z}_u, \mathbf{z}_v) + \omega\mathcal{L}_\text{Uniform}(\mathbf{z}_u),
\end{equation}
where $\omega$ is the coefficient that balances alignment and uniformity. 
Notably, we remove the uniformity optimization for item representations, further insights are presented in Section \ref{sec:discuss}. 

\subsubsection{\textbf{Representation Matching}}
Although the preceding objectives optimize overall representations, discrepancies remain across spaces \cite{Peng_TOIS_DALR_2025}. 
In particular, collaborative and intent spaces show inconsistencies in distribution, geometry, and coverage, which may lead to representation degradation \cite{zhang_TKDE_CoLLM_2025,CARec_CIKM_2024}. 
To address this, we propose two matching approaches.

\noindent\textbf{Coarse-grained Matching.} 
Although the collaborative alignment objective partially improves representation quality, discrepancies across different spaces still exist \cite{zhang_SIGIR_DMRec_2025}. Specifically, reconstructed collaborative representations often deviate from their semantic or intent counterparts, resulting in distributional gaps on the shared hypersphere. 
To mitigate this problem, coarse-grained matching anchors each representation to its corresponding semantic or intent representation, thereby promoting global consistency across spaces. Formally, coarse-grained matching loss is defined as follow: 
\begin{equation}
\label{eq:coarse_match}
\mathcal{L}_{\text{Coarse}}
= \mathbb{E}_{x\in\mathcal B_x}
\;\underbrace{ \mathrm{max}(1-\cos(\mathbf z_x,\ \mathbf W\,\gamma_x)}_{\text{Coarse Matching}})
\\ + \underbrace{\|\mathbf W^\top\mathbf W-\mathbf I\|_2^2}_{\text{Constraint}},
\end{equation}
where $\gamma_x \in \{\mathbf s_x,\mathbf c_x^{\mathrm{Pro}}\}$, and $\cos(\cdot,\cdot)$ denotes cosine similarity. 
Here, An orthogonal mapping $\mathbf W \in \mathbb{R}^{d\times d}$ is introduced to project semantic signal into the dual-intent space for alignment, while the orthogonality constraint regularizes $\mathbf W$ to preserve geometric consistency across spaces \cite{lample_ICLR_2018}.

\noindent\textbf{Fine-grained Matching.} 
On the hypersphere, user and item representations are ideally expected to be uniformly distributed to support global optimization. 
In practice, however, some samples become isolated, drifting away from the overall distribution \cite{wang_ICML_understand_2020}. 
To address isolation, fine-grained matching first mines a nearest neighbor set for all representations and then aligns each representation to a similarity-selected sample. 
Formally, the nearest-neighbor set is defined as follow:
 \begin{equation}
\mathcal{J}^+ 
= \{j_i^*=\arg\max_{j \in (\mathcal{U};\mathcal{V})\setminus\{i\}} \mathrm{Sim}(\mathbf z_i,\mathbf z_j)\},\; i \in (\mathcal{U};\mathcal{V}),
\end{equation}
where $\mathcal{J}^+$ denote the set of retrieved candidate. 
The similar positive pairs $(\mathbf z_i,\mathbf c^{\text{Pro}}_{j_i^*})$ are defined from empirical joint distribution $\mathbb{P}(\mathbf z,\mathbf c)$, 
while mismatched pairs $(\mathbf z_i,\mathbf c^{\text{Pro}}_{\setminus j_i^*})$ are sampled independently from marginals $\mathbb{P}(\mathbf z)\mathbb{P}(\mathbf c)$. 
Inspired by the mutual information (MI) \cite{poole_ICLR_MI_2019} theory, 
we define fine-grained matching loss as follow:
\begin{equation}
\label{eq:fine_match}
\mathcal L_{\text{Fine}}
= 
\mathbb E_{\mathbb{P}(\mathbf{z},\mathbf{c})}
\underbrace{\cos(\mathbf z_i,\mathbf c^{\text{Pro}}_{j^*_i})}_{\text{Fine Matching}}
\;-\;
\log\,
\mathbb E_{ \mathbb P(\mathbf z)\mathbb P(\mathbf c)}
\underbrace{\varphi(\cos(\mathbf z_i,\mathbf c^{\text{Pro}}_{\setminus j_i^*}))}_{\text{Normalization}}
.
\end{equation}  
By explicitly reinforcing local similarity, fine-grained matching effectively mitigates sample isolation, thereby enhancing the structural consistency of representations on the hypersphere.

\subsubsection{\textbf{Representation Regularization}}
In addition, to further consolidate representations, we incorporate self-supervised learning (SSL) \cite{liu_TKDE_graph_2022,gui_TPAMI_survey_2024} as the regularization technique. 
The core paradigm of SSL, contrastive learning (CL), aims to maximize the consistency between positive samples while simultaneously pushing them away from negatives. The widely adopted contrastive loss \cite{oord_Info_2018} is formulated as follows: 
\begin{equation}
\mathcal{L}_\text{Reg}(\mathbf{e}_x,\mathbf{e}_{x'}) 
= \varphi(\mathbf{e}_x^{\top}\mathbf{e}_{x'}/\tau)/\sum_{x'' \in \mathcal{B}}\varphi(\mathbf{e}_x^{\top}\mathbf{e}_{x''}/\tau),
\end{equation}
where $\tau$ is the temperature hyperparameter, and $\mathbf{e}_x$ and $\mathbf{e}_{x'}$ denote the diverse regularization views.  

\noindent\textbf{Intra-space Regularization.} 
Representation within the hypersphere space may suffer from collapse or redundancy \cite{chen_WSDM_collapse_2024}, which weakens their expressiveness and discriminability. 
To mitigate this, intra-space regularization applies contrastive constraints within collaborative representations, encouraging users and items to maintain sufficient diversity while avoiding degenerate distributions: 
\begin{equation}
\label{eq:intra_reg}
\mathcal{L}_\text{Intra} = \mathcal{L}_\text{Reg}(\mathbf{z}_u,\mathbf{z}_{u,\mu}) + \mathcal{L}_\text{Reg}(\mathbf{z}_v,\mathbf{z}_{v,\mu}).
\end{equation}

\noindent\textbf{Interaction Regularization.} 
Beyond intra-space consistency, recommendation fundamentally relies on capturing collaborative signals from interaction data. 
Interaction regularization enforces closeness between users and their positive items, while simultaneously distinguishing them from negatives. 
Specifically, we apply the contrastive loss to both the final embeddings and the mean representations of users and items:
\begin{equation}
\label{eq:inter_reg}
\mathcal{L}_\text{Inter} = \mathcal{L}_\text{Reg}(\mathbf{z}_u,\mathbf{z}_v) + \mathcal{L}_\text{Reg}(\mathbf{z}_{u,\mu},\mathbf{z}_{v,\mu}).
\end{equation}

\subsubsection{\textbf{Multi-task Joint Optimization}}
To jointly optimize recommendation with representation spaces, we adopt a multi-task learning framework \cite{wu_SIGIR_SGL_2021,SimGCL_yu_2022_SIGIR}. 
The main task is guided by the alignment–uniformity loss $\mathcal{L}_\text{AU}$, while additional objectives from matching and regularization provide complementary signals. 
The overall optimization objective is defined as:
\begin{equation}
\label{eq:total_loss}
\mathcal{L}_\text{Rec} = \mathcal{L}_\text{AU} 
+ \lambda_1 (\mathcal{L}_\text{Coarse}  
+ \mathcal{L}_\text{Fine}) 
+ \lambda_2 ( \mathcal{L}_\text{Intra} 
+  \mathcal{L}_\text{Inter} )
+ \|\Theta\|_2^2,
\end{equation}
where $\lambda_1,\lambda_2$ are trade-off hyperparameters, and $\Theta$ are the trainable model parameters. 
The complete training process of DIAURec is presented in Algorithm \ref{alg:diaurec}.

\begin{algorithm}[t]
\caption{Training Process of DIAURec}
\label{alg:diaurec}
\begin{flushleft}
\textbf{Input:} 
user--item interaction set $\mathcal E$ (or matrix $\mathbf R$); 
multi-source textual information $\{T_k(u;v)\}$; 
LLM profile generator $\mathcal F_{\text{LLM}}(\cdot)$; 
number of intents $\mathcal K$;  
learning rate $\eta$. \\
\textbf{Output:} trained model parameters $\Theta$.
\end{flushleft}
\begin{algorithmic}[1]
\State Initialize and update base representations $\mathbf z_{u,\mu},\mathbf z_{v,\mu}$ (Eq. \ref{eq:col_model});
\State Pre-compute profiles $\mathcal P(u;v)$ via $\mathcal F_{\text{LLM}}(T_k(u;v))$ and obtain semantic representations $\mathbf s_u,\mathbf s_v$ (Eq. \ref{eq:sem_model});

\While{DIAURec not converged}
    \State Sample a mini-batch of positive interactions $\mathcal B \subset \mathcal E$;
    \For{$\langle u,v\rangle \in \mathcal B$}

        \LongState{Model prototype intents $\mathbf c_u^{\mathrm{Pro}},\mathbf c_v^{\mathrm{Pro}}$ from $\mathbf s_u,\mathbf s_v$ (Eq. \ref{eq:pro_intent_model});}
        
        \LongState{Sample distribution directions $\mathbf h_u^{\mathrm{Dis}},\mathbf h_v^{\mathrm{Dis}}$ via vMF and model distribution intents $\mathbf c_u^{\mathrm{Dis}},\mathbf c_v^{\mathrm{Dis}}$ from $\mathbf z_{u,\mu},\mathbf z_{v,\mu}$  
        (Eqs. \ref{eq:map_distribution}--\ref{eq:dis_intent_model});}

        \State Reconstruct representations $\mathbf z_u,\mathbf z_v$ (Eqs. \ref{eq:reconstruction_gcn}--\ref{eq:reconstruction_overall});
        
    \EndFor
    \LongState{\textit{// Batch-level representation optimization on $\mathcal B$.}}
     \LongState{Calculate representation learning ($i.e.,$ alignment and uniformity) loss $\mathcal L_{\mathrm{AU}}$ (Eq. \ref{eq:item_uniform});}
        \State Calculate coarse matching loss $\mathcal L_{\mathrm{Coarse}}$ (Eq. \ref{eq:coarse_match});
        \State Calculate fine matching loss $\mathcal L_{\mathrm{Fine}}$ (Eq. \ref{eq:fine_match});
        \LongState{Calculate regularization losses $\mathcal L_{\mathrm{Intra}},\mathcal L_{\mathrm{Inter}}$ (Eqs. \ref{eq:intra_reg}--\ref{eq:inter_reg});}
        
        \LongState{Aggregate total losses to form the joint optimization objective $\mathcal L_{\mathrm{Rec}}$
        (Eq. \ref{eq:total_loss});}
    \LongState{Average gradients from mini-batch and update parameters:
    $\Theta \leftarrow \Theta - \eta \nabla_{\Theta}\mathcal L_{\mathrm{Rec}}$;}
\EndWhile
\State \Return $\Theta$;
\end{algorithmic}
\end{algorithm}

\subsection{Discussion}
\subsubsection{\textbf{Time Complexity}}
We analyze the computational complexity of DIAURec and compare it with existing recommendation models.
Let $|\mathcal{E}|$, $|\mathcal{U}|$, and $|\mathcal{V}|$ denote the numbers of interactions, users, and items, respectively.
Let $d$ be the embedding dimension, $L$ the number of GCN layers, and $\mathcal K$ the number of intent prototypes.

The dominant cost of the graph encoder is $\mathcal O(L|\mathcal{E}|d)$, which is identical to commonly used GCN-based CF models.
The dual-intent modeling module introduces an additional cost of $\mathcal O((|\mathcal{U}|+|\mathcal{V}|)\mathcal Kd)$ for learning intent-aware representations. During mini-batch training with batch size $\mathcal B$, the alignment and uniformity objectives incur a cost of $\mathcal O(\mathcal Bd + \mathcal B^2 d)$. Similarly, the coarse- and fine-grained matching as well as intra-space and interaction regularization terms are computed with the same order of complexity $\mathcal O(\mathcal Bd + \mathcal B^2 d)$. 
Despite introducing dual-intent modeling and multi-level alignment objectives, DIAURec maintains a comparable computational complexity to existing methods, while enabling richer representation interactions.

\subsubsection{\textbf{Relation with LLM-based Recommendation Methods}}
As discussed in the Introduction, recent LLM-based methods \cite{bao_RS_TALLRec_2023,RLMRec_2024_WWW} have explored incorporating large language models to enhance semantic understanding.
From a methodological perspective, existing studies can be broadly divided into two paradigms.

The first paradigm tightly integrates LLMs into the recommendation pipeline, such as through model fine-tuning, prompt-based reasoning, or chain-of-thought inference. 
These methods leverage the generative and reasoning capabilities of LLMs to directly guide recommendation decisions. 
However, they inevitably introduce substantial computational overhead in both training and inference, due to repeated LLM invocation and large-scale parameter optimization. 
The second paradigm employs LLMs as language encoders to obtain enriched semantic representations, and focuses on how to incorporate such representations into CF. 
Nevertheless, existing methods in this paradigm mainly emphasize representation extraction \cite{RLMRec_2024_WWW}, and increasingly rely on sophisticated prompt designs or multi-stage semantic modeling \cite{2025IRLLRec}, which can be computationally expensive and often leave the interaction between semantic and collaborative spaces underexplored. 

DIAURec follows the second paradigm. 
However, rather than pursuing more complex semantic generation or LLM-based reasoning, our focus is on how to learn, exploit, and optimize the  representation space within CF. 
Specifically, we study how coarse- and fine-grained semantic representations can be structured into a dual-intent space, and how multi-level objectives can be designed to jointly match diverse representation spaces. 
From this perspective, DIAURec can be viewed as a representation optimization framework, which shifts the emphasis from acquiring richer semantic features to modeling their interactions with collaborative signals.

\subsubsection{\textbf{Relation with Uniformity}}
\label{sec:discuss}

Recently, researchers have introduced \emph{uniformity} to characterize representations \cite{SimGCL_yu_2022_SIGIR,gao_simcse_2021}. 
To further explore this property, we consider the representation $\mathbf z_m \in \mathcal{B}$ with the naive uniformity loss $\mathcal{L}_{\text{Uniform}}(\mathbf{z})$ from Eq. \ref{eq:uniform}. 

Let $d_{mn}^2=\|\mathbf z_m-\mathbf z_n\|_2^2$ and $S=\sum_{a\neq b}\varphi(-d_{ab}^2)$. 
For each representation $\mathbf z_m$, the gradient can be derived as follows: 
\begin{align}
\label{eq:uni_grad}
\frac{\partial \mathcal L_{\mathrm{Uniform}}(\mathbf{z}_m)}{\partial \mathbf z_m}
&= \tfrac{1}{S}\sum_{m\neq n}\varphi(-d_{mn}^2)\left(-\frac{\partial d_{mn}^2}{\partial \mathbf z_m}\right)
\nonumber \\
&= -4\sum_{m\neq n}\pi_{mn}\,(\mathbf z_m-\mathbf z_n),
\end{align}
where the normalized weights $\pi_{mn}$ are defined as follows:
\begin{equation}
\pi_{mn}=\varphi(-d_{mn}^2)/S, \qquad m\neq n.
\end{equation}
The above weights assign greater contribution to sample pairs with high similarity, 
which makes the gradient primarily sensitive to nearest neighbors. 
Consequently, the update acts to separate highly similar representation, expanding dense regions and encouraging a more uniform coverage on the hypersphere. 

To further illustrate this mechanism, the gradient can be reformulated in matrix form as follows:
\begin{equation}
\frac{\partial \mathcal L_{\mathrm{Uniform}}(\mathbf{Z})}{\partial \mathbf{Z}} = -4\mathbf{LZ},
\end{equation}
where $\mathbf{Z}\in\mathbb{R}^{\mathcal{B}\times d}$ , $\mathbf{L}=\mathbf{D}-\mathbf{W}$ is the weighted graph Laplacian \cite{belkin_laplacian_2003}, and $\mathbf{W}=[w_{mn}]$ with $w_{mn}=\pi_{mn}$. 
Equivalently, uniformity can be interpreted through the quadratic form \cite{belkin2006quadratic}: 
\begin{equation}
\mathrm{tr}(\mathbf{Z}^\top \mathbf{L}\mathbf{Z})
= \tfrac{1}{2}\sum_{m\neq n}w_{mn}\|\mathbf z_m-\mathbf z_n\|_2^2.
\end{equation}
We argue that uniformity increases the Laplacian energy of the representation graph, thereby acting as an \emph{anti-smoothing} operator that pushes representations away from the weighted averages of their neighbors. 
For user representations, this repulsion alleviates over-crowding and encourages broader coverage of diverse intents. 
However, for item representations, nearest neighbors are often highly semantically coherent. 
Forcing them apart tends to disrupt meaningful clusters and weaken fine-grained similarity, making it difficult for the model to capture items that reflect user preferences. Although such treatment may improve the overall uniformity, its side effects are non-negligible. 
We therefore introduce the regularization technologies to mitigate this conflict \cite{SimGCL_yu_2022_SIGIR}. 
This explains why enforcing consistency is beneficial for user representations but may have adverse effects on item representations.

\section{Experiments}

\begin{table}
  \caption{Statistics of the datasets.}
  \label{tab:dataset_statistics}
  \begin{tabular}{c|c|c|c|c}
    \bottomrule
    \textbf{Dataset} & \textbf{\#Users} & \textbf{\#Items} & \textbf{\#Interactions} & \textbf{Sparsity} \\
    \hline
    \hline
    Amazon-book & 11,000 & 9,331 & 200,860 & 99.81\% \\
    \hline
    Yelp & 11,091 & 11,010 & 166,620 & 99.86\%  \\
    \hline
    Steam & 23,310 & 5,236 & 525,923 & 99.57\%  \\
    \toprule
  \end{tabular}
\end{table}

\renewcommand{\arraystretch}{1.2}
\begin{table*}
    \centering
    \footnotesize
\caption{A comparison between DIAURec and representative state-of-the-art baselines is reported. The best results are highlighted in bold, and the second-best are underlined. R@ and N@ denote Recall@ and NDCG@, respectively. “Improv.\%” represents the relative improvement over the strongest baseline. Statistical significance is assessed using a t-test with $p < 0.05$.
}
    \begin{tabular}{p{1.9cm}|p{0.505cm}p{0.505cm}p{0.505cm}p{0.505cm}p{0.505cm}p{0.55cm}|p{0.505cm}p{0.505cm}p{0.505cm}p{0.505cm}p{0.505cm}p{0.55cm}|p{0.505cm}p{0.505cm}p{0.505cm}p{0.505cm}p{0.505cm}p{0.55cm}}
    \hline
    \multicolumn{1}{l|}{\textbf{Dataset}} & \multicolumn{6}{c|}{\textbf{Amazon-book}} &
    \multicolumn{6}{c|}{\textbf{Yelp}} & \multicolumn{6}{c}{\textbf{Steam}} \\
    \cline{0-18} 
     Method  & R@5 & R@10 & R@20 & N@5 & N@10 & N@20 & R@5 & R@10 & R@20 & N@5 & N@10 & N@20 & R@5 & R@10 & R@20 & N@5 & N@10 & N@20 \\
    \hline
    \hline
    LightGCN(SIGIR'20)
    &0.0570 &0.0915  &0.1411  &0.0537 &0.0694  & 0.0856 
    &0.0421 &0.0706  &0.1157  &0.0491 &0.0580  & 0.0733
    &0.0518 &0.0852  &0.1348  &0.0575 &0.0687  & 0.0855
    \\
    SGL-ED(SIGIR'21)
    &0.0637  &0.0994  &0.1473  &0.0632 &0.0756  & 0.0913
    &0.0432  &0.0722  &0.1197  &0.0501 &0.0592  & 0.0753
    &0.0562  &0.0915  &0.1430  &0.0616 &0.0731  & 0.0913
    \\
    SimGCL(SIGIR'22)
    &0.0618 &0.0992  &0.1512  &0.0619 &0.0749  & 0.0919 
    &0.0467 &0.0772  &0.1254  &\underline{0.0546} &0.0638  & 0.0801
    &0.0564 &0.0917  &0.1436  &0.0618 &0.0736  & 0.0915\\
    DCCF(SIGIR'23)
    &\underline{0.0652} &\underline{0.1019}  &0.1517  &\underline{0.0654} &\underline{0.0780}  &\underline{0.0943} 
    &\underline{0.0468} &0.0778  &0.1249  &0.0543 &0.0640  & 0.0800
    &0.0561 &0.0915  &0.1437  &0.0618 &0.0735  & 0.0914\\
    BIGCF(SIGIR'24)
    &0.0644 &0.1000  &0.1496  &0.0652 &0.0767  & 0.0932 
    &0.0458 &0.0758  &0.1237  &0.0536 &0.0627  &0.0789	
    &0.0560 &0.0901  &0.1431  &0.0612 &0.0725  & 0.0905 \\
    \hline
    KAR(RecSys'24)
    &0.0596 &0.0934  &0.1416  &0.0590 &0.0705  & 0.0860 
    &0.0437 &0.0740  &0.1194  &0.0506 &0.0602  & 0.0756
    &0.0519 &0.0854  &0.1353  &0.0578 &0.0690 & 0.0854\\
    LLMRec(WSDM'24)
    &0.0605 &0.0963  &0.1469  &0.0600 &0.0715  & 0.0855 
    &0.0439 &0.0744  &0.1203  &0.0509 &0.0605  & 0.0751
    &0.0543 &0.0901  &0.1431  &0.0604 &0.0722  & 0.0901\\
    RLMRec(WWW'24)
    &0.0608 &0.0969  &0.1483  &0.0606 &0.0734  & 0.0903 
    &0.0445 &0.0754  &0.1230  &0.0518 &0.0614  & 0.0776
    &0.0548 &0.0907  &0.1433  &0.0608 &0.0729  & 0.0907\\
    AlphaRec(ICLR'25)
    &0.0595 &0.0930  &0.1412  &0.0597 &0.0708  & 0.0873 
    &0.0439 &0.0748  &0.1213  &0.0507 &0.0603  & 0.0752
    &0.0538 &0.0897  &0.1420  &0.0600 &0.0718  & 0.0898\\
    IRLLRec(SIGIR'25)
    &0.0643 &0.1009  &\underline{0.1538}  &0.0638 &0.0765  & 0.0938 
    &0.0460 &\underline{0.0781}  &\underline{0.1278}  &0.0542 &\underline{0.0642}  & \underline{0.0810}
    &\underline{0.0566} &\underline{0.0918}  &\underline{0.1445}  &\underline{0.0620} &\underline{0.0737}  & \underline{0.0919}\\
    \hline
    DirectAU(KDD'22)
    &0.0588 &0.0920 &0.1382 &0.0585 &0.702 &0.0854	
    &0.0436 &0.0736 &0.1200 &0.0513 &0.0606 &0.0763
    &0.0493 &0.0991 &0.1249 &0.0540 &0.0641 &0.0796 \\
    MAWU(CIKM'23)
    &0.0606 &0.0924  &0.1392  &0.0591 &0.0704  & 0.0860
    &0.0425 &0.0735  &0.1204  &0.0496 &0.0593  & 0.0751
    &0.0460 &0.0752  &0.1185  &0.0497 &0.0596  & 0.0746\\
    GraphAU(CIKM'23)
    &0.0581 &0.0924  &0.1400  &0.0549 &0.0673  & 0.0832 
    &0.0438 &0.0745  &0.1226  &0.0523 &0.0618  & 0.0779
    &0.0555 &0.0910  &0.1434  &0.0598 &0.0718  & 0.0897\\
    CARec(CIKM'24)
    &0.0579 &0.0931  &0.1392  &0.0583 &0.0702  & 0.0854 
    &0.0422 &0.0700  &0.1130  &0.0483 &0.0568  & 0.0714
    &0.0562 &0.0900  &0.1384  &0.0598 &0.0713  & 0.0880\\
    SIURec(AAAI'25)
    &0.0572 &0.0922 &0.1378 &0.0553 &0.0688 &0.0832
    &0.0427 &0.0712 &0.1175 &0.0492 &0.0588 &0.0743
    &0.0467 &0.0771 &0.1239 &0.0503 &0.0613 &0.0773\\
    \rowcolor[gray]{0.9}
    \textbf{DIAURec (Ours)}
    &\textbf{0.0744} &\textbf{0.1155} &\textbf{0.1701} 
    &\textbf{0.0745} &\textbf{0.0886} &\textbf{0.1063}
    &\textbf{0.0488} &\textbf{0.0809} &\textbf{0.1332} 
    &\textbf{0.0571} &\textbf{0.0669} &\textbf{0.0847}
    &\textbf{0.0606} &\textbf{0.0978} &\textbf{0.1526} 
    &\textbf{0.0660} &\textbf{0.0784} &\textbf{0.0971} 
    \\
    \hline
    \hline
    Improv.\% 
    &14.11\% &10.60\% &10.60\% &13.91\% &13.59\% &12.73\%
    &4.27\% &3.59\% &4.23\% &4.58\% &4.21\% &4.57\%
    &7.07\% &6.54\% &5.61\% &6.45\% &6.38\% &5.66\% \\
    $p$-value
    &\smash{1.29e-8} &\smash{3.37e-8} &\smash{3.22e-10}
    &\smash{5.07e-8} &\smash{2.03e-8} &\smash{2.14e-9}
    &\smash{3.62e-7} &\smash{4.66e-8} &\smash{1.79e-7}
    &\smash{8.98e-9} &\smash{7.59e-9} &\smash{7.46e-9}
    &\smash{4.89e-10} &\smash{5.33e-8} &\smash{6.01e-10}
    &\smash{8.94e-9} &\smash{6.31e-8} &\smash{3.19e-9} \\
    \hline
    \end{tabular}
        
    \label{tab:main}
\end{table*}

We evaluate our method with the following research questions: 
 
\begin{itemize}[leftmargin=*,label=--]
    \item \textbf{RQ1:} How does DIAURec perform compared with state-of-the-art methods?  
    \item \textbf{RQ2:} Do the core modules of DIAURec contribute significantly to performance improvements?  
    \item \textbf{RQ3:} In what ways do different optimization objectives influence the effectiveness of DIAURec? 
    \item \textbf{RQ4:} What is the impact of the encoder design and the number of GCN layers on the performance of DIAURec? 
    \item \textbf{RQ5:} How sensitive is the model to key hyperparameters?  
\end{itemize}

\subsection{Experimental Settings}
\subsubsection{\textbf{Datasets.}} 
We conduct experiments on three benchmark datasets \cite{RLMRec_2024_WWW,2025IRLLRec}: Amazon-book, Yelp, and Steam. 
For Amazon-book and Yelp, only interactions with ratings no less than 3 are retained, while Steam is used directly due to the absence of rating scores. 
Subsequently, $k$-core filtering is applied to remove inactive users and items, and the largest connected component is preserved \cite{DCCF_2023_SIGIR}. 
Each dataset is split into training, validation, and test sets with a ratio of 3:1:1. 
The statistics are summarized in Table~\ref{tab:dataset_statistics}.   

\subsubsection{\textbf{Baselines.}} 
We validate the effectiveness of DIAURec with three categories of representative state-of-the-art methods:  
\begin{itemize}[leftmargin=*,label=--]
    \item \textbf{ID-based Methods:} LightGCN \cite{he_LightGCN_SIGIR_2020}, SGL-ED \cite{wu_SIGIR_SGL_2021}, SimGCL \cite{SimGCL_yu_2022_SIGIR}, DCCF \cite{DCCF_2023_SIGIR}, and BIGCF \cite{zhang_BIGCF_SIGIR_2024}.
    \item \textbf{LLM-based Methods:} KAR \cite{KAR_2024_RS}, LLMRec \cite{LLMRec_2024_WSDM}, RLMRec \cite{RLMRec_2024_WWW}, AlphaRec \cite{AlphaRec_2025_ICLR}, and IRLLRec \cite{2025IRLLRec}.
    \item \textbf{AU-based Methods:} DirectAU \cite{DirectAU_wang_2022_KDD}, MAWU \cite{MAWU_Park_2023_CIKM}, GraphAU \cite{GraphAU_Yang_2023_CIKM}, CARec \cite{CARec_CIKM_2024}, and SIURec \cite{SIURec}.
\end{itemize}
\subsubsection{\textbf{Hyperparameter Settings.}} 
We implement DIAURec in PyTorch \cite{Pytorch_NIPS}. 
For fair comparison, the embedding and batch size are set to 32 and 4096, respectively, across all methods. 
The number of GCN layers for graph-based methods is fixed at 3. 
For all AU-based baselines, we set the weight $\omega$ = 1.0 on Amaznon-book dataset, and $\omega$ = 0.4 on Yelp and Steam datasets. 
We adopt Adam \cite{kingma2014adam} as the optimizer with an initial learning rate of $1e^{-4}$. 
The temperature coefficient $\tau$ is fixed at 0.2. 
The intent coefficient $\eta$ and intent representation size are fixed at 1.0 and 128, respectively.  
The $\lambda_{1} $ and $ \lambda_{2}$ are tuned from the range of $[0.1,0.15,0.2,0.25,0.3]$.
Early stopping \cite{prechelt2002early} is triggered if the validation performance does not improve for 5 consecutive epochs. 
Evaluation is conducted using Recall@$N$ and NDCG@$N$ ($N=5,10,20$) \cite{he_LightGCN_SIGIR_2020,wang_SIGIR_NGCF_2019}, and all reported results are averaged over five independent runs.  

\subsection{Performance Comparison (RQ1)}

\subsubsection{\textbf{Comparison with Baselines}}

Table~\ref{tab:main} reports the results of DIAURec and all baselines on three datasets. DIAURec achieves the best performance across the three categories of baselines. In terms of Recall@20, DIAURec surpasses the strongest baselines by 10.31\%, 4.23\%, and 5.61\% on Amazon-Book, Yelp, and Steam, respectively. These consistent improvements demonstrate the superiority of our method under different recommendation scenarios. We attribute the performance gains to the proposed representation modeling and optimization modules, which enable effective optimization across different spaces and alleviate representation collapse as well as the inconsistency between semantic and collaborative spaces. 

IRLLRec \cite{2025IRLLRec}, as a method that leverages LLM to mine user intents, achieves superior recommendation performance compared with RLMRec \cite{RLMRec_2024_WWW} and AlphaRec \cite{AlphaRec_2025_ICLR}, highlighting the advantage of intent modeling. However, IRLLRec only designs finer-grained representation construction without effective optimization mechanisms, which not only increases the burden of the representation space but also introduces semantic noise that leads to biased recommendations. Therefore, compared with all LLM-based methods, DIAURec achieves better results, mainly due to its explicit representation optimization strategy. On the other hand, compared with AU-based methods, although some methods ($e.g.,$ CARec \cite{CARec_CIKM_2024}) also incorporate semantic information, their lack of intent-aware modeling and failure to achieve joint optimization across different representation spaces result in inferior performance to DIAURec.  
   
\subsubsection{\textbf{Comparison with Data Sparsity}}
\begin{figure}[!t]
  \centering
\raisebox{2.25cm}{\rotatebox[origin=c]{90}{\small \textbf{NDCG@20(\%)}}}
  \hfill
  \begin{subfigure}[b]{0.31\linewidth}
    \includegraphics[width=\linewidth]{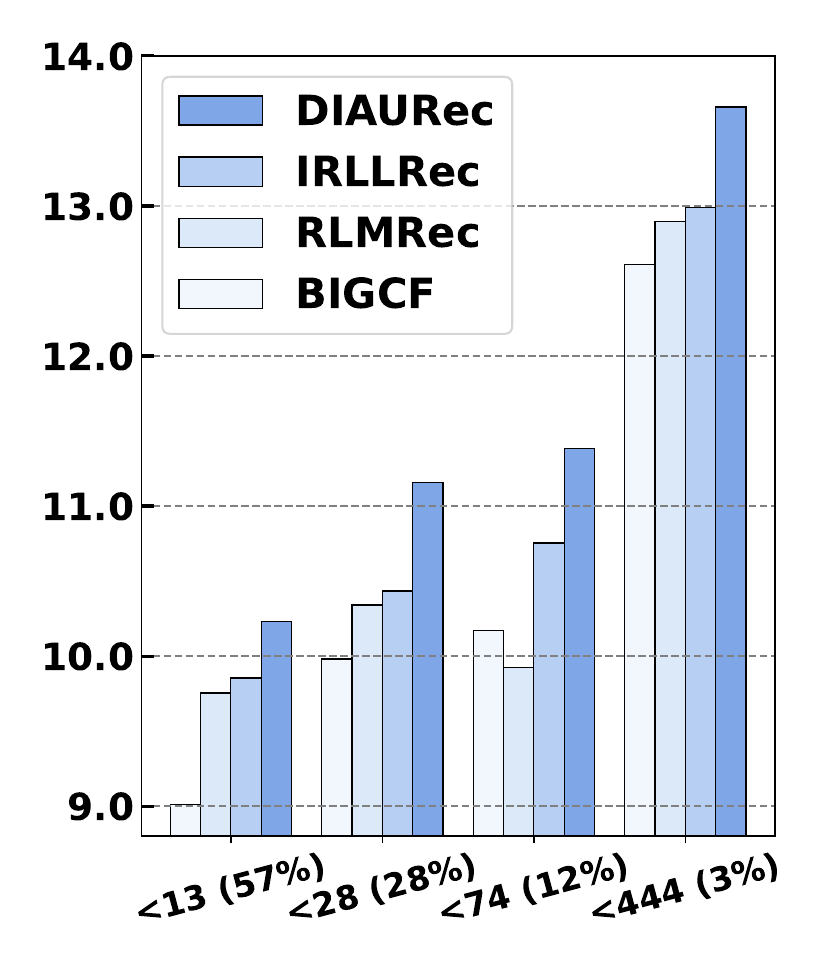}
    \caption{Amazon-book}
    \label{fig:yelp2018_sparsity_ndcg@20}
  \end{subfigure}
  \hfill 
  \begin{subfigure}[b]{0.31\linewidth}
    \includegraphics[width=\linewidth]{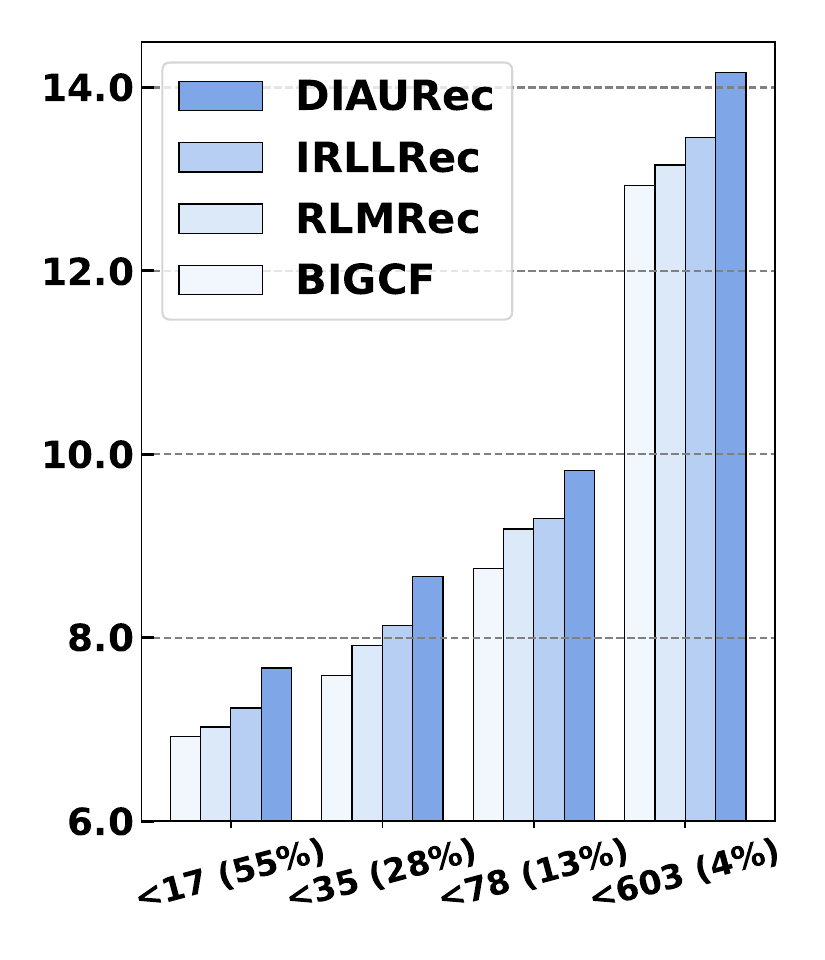}
    \caption{Yelp}
    \label{fig:tmall_sparsity_ndcg@20}
  \end{subfigure}
  \hfill 
  \begin{subfigure}[b]{0.31\linewidth}
    \includegraphics[width=\linewidth]{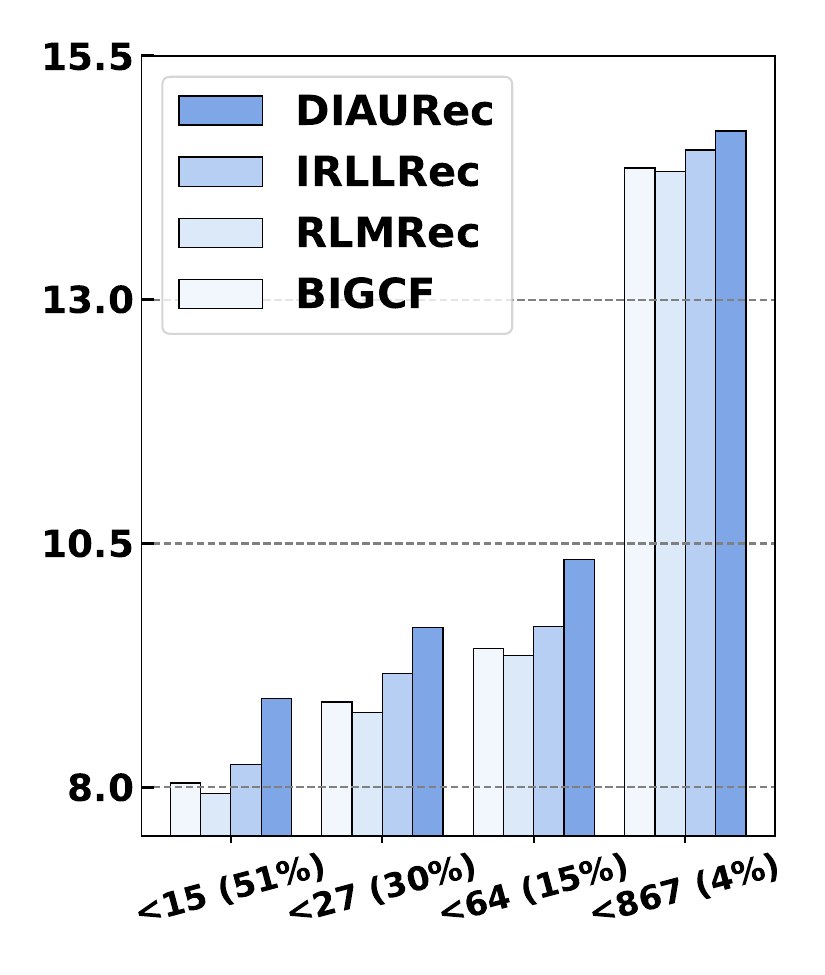}
    \caption{Steam}
    \label{fig:tmall_sparsity_ndcg@20}
  \end{subfigure}
  \caption{Sparsity study of DIAURec with several baselines \textit{w.r.t.} NDCG@20 on Amazon-book, Yelp, and Steam datasets. The x-axis denotes diverse user proportions.}
  \label{fig:sparsity}
\end{figure}

Here, we investigate the performance of DIAURec under data sparsity, using representative baselines including the LLM-based RLMRec and the intent-based BIGCF and IRLLRec for comparison. Specifically, we divide the training set into four user groups according to the number of interactions (as shown on the x-axis of Fig. \ref{fig:sparsity}). The leftmost group contains users with the fewest interactions, representing the sparsest scenario, while the rightmost group corresponds to the most active users. All models are trained on the complete training set and evaluated across these sparsity groups. The experimental results show that DIAURec achieves significant improvements across all groups, with relative gains of 4\%, 6\%, and 8\% on the sparsest group of the three datasets, respectively. These advantages mainly stem from the proposed prototype and distribution intent modeling strategy, which not only enriches collaborative representations under limited interactions but also introduces additional supervision signals for the recommendation task through coarse- and fine-grained matching together with dual regularization, thereby enhancing the model’s resistance to sparsity.

\begin{figure}[!t]
  \centering
  \hfill
  \begin{subfigure}[b]{0.49\linewidth}
    \includegraphics[width=\linewidth]{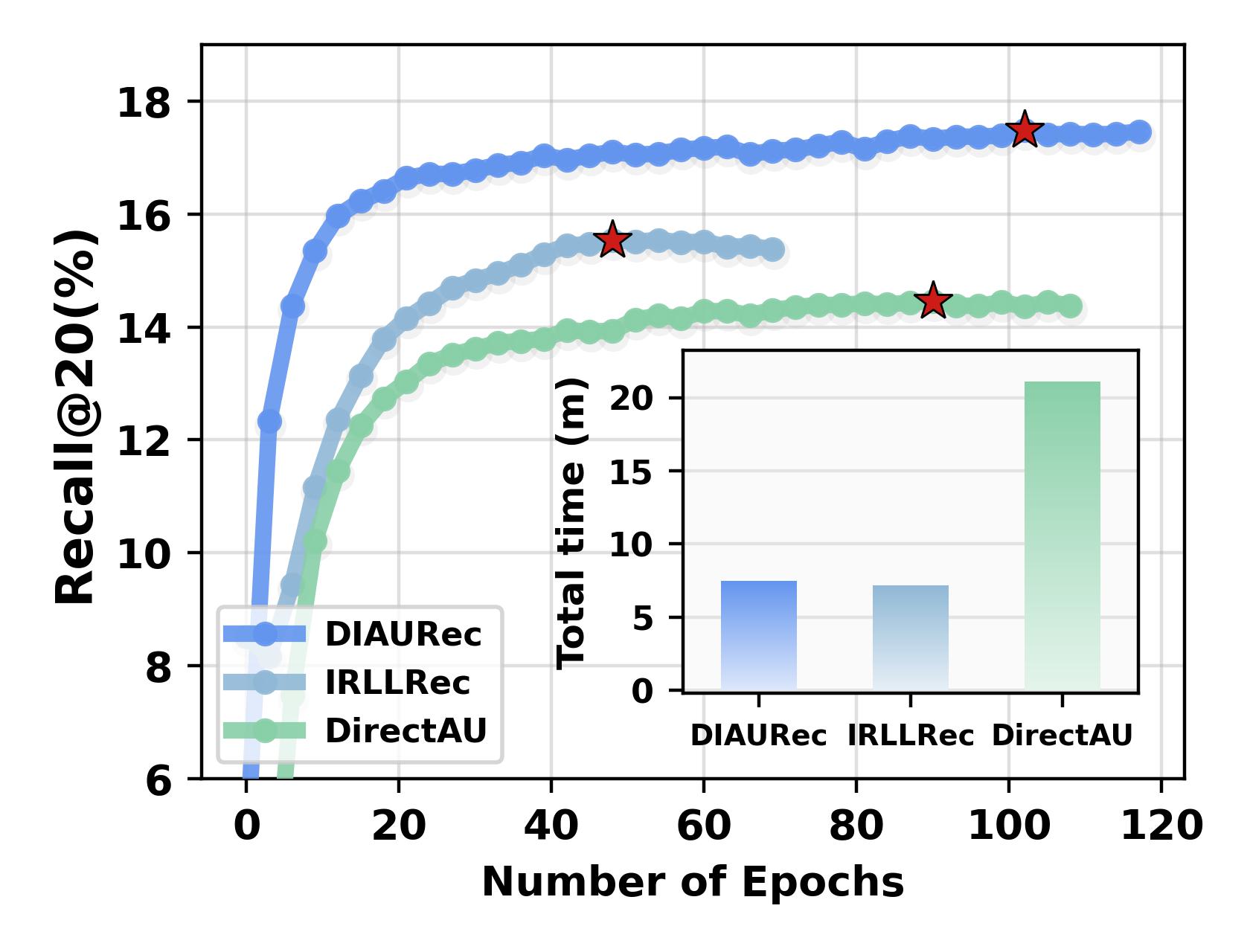}
    \caption{Amazon-book}
  \end{subfigure}
  \hfill 
  \begin{subfigure}[b]{0.49\linewidth}
    \includegraphics[width=\linewidth]{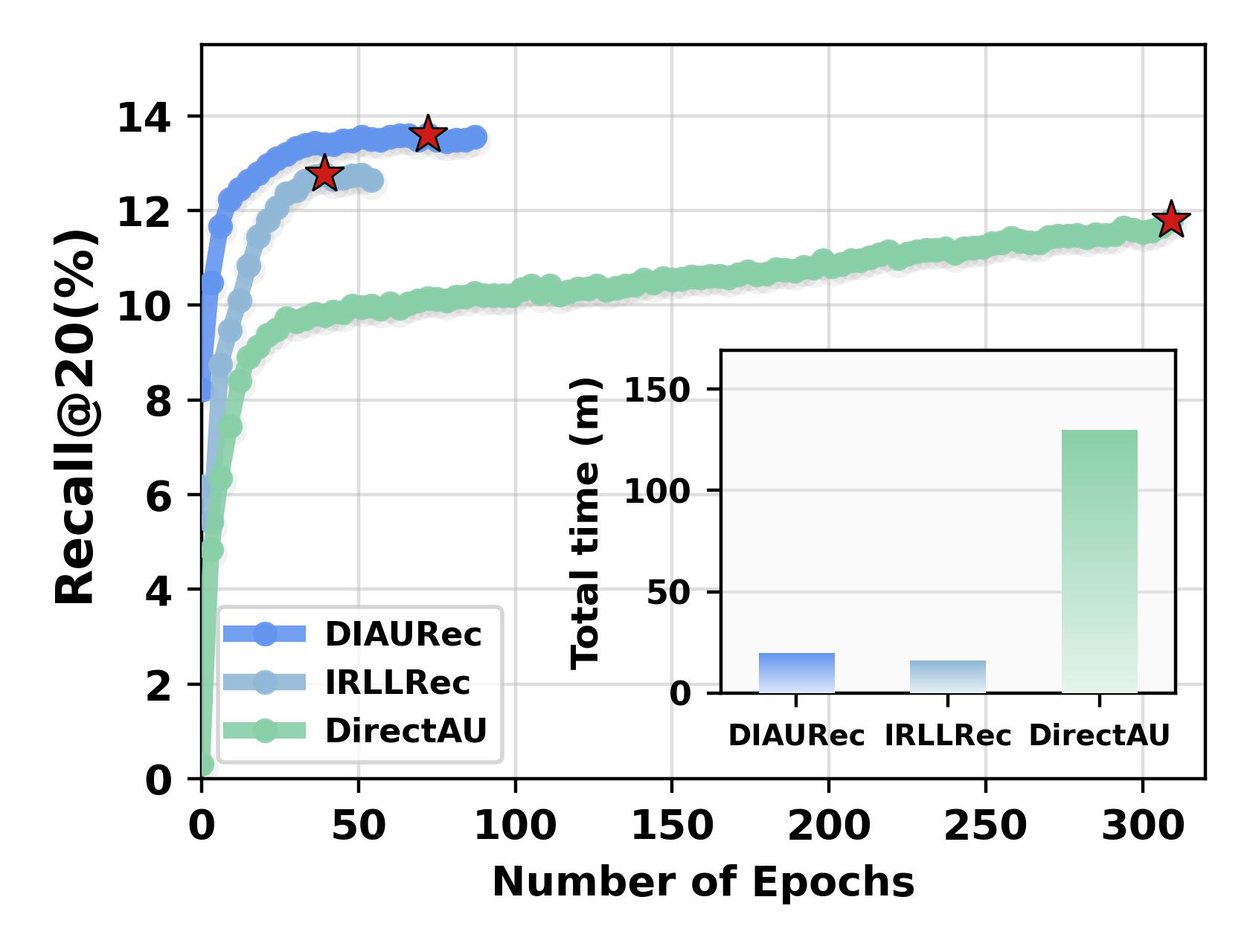}
    \caption{Yelp}
  \end{subfigure}
  \caption{Comparison \textit{w.r.t.} Recall@20 during training process and total time for DIAURec and several baselines across the Amazon-book and Yelp datasets. (m: minutes)}
  \label{fig:train_time}
\end{figure}

\subsubsection{\textbf{Comparison with Training Efficiency}}
To further demonstrate the training efficiency of DIAURec, we report the training curves and total training time of different methods in Figure \ref{fig:train_time}. We observe that, as a representative AU-based method, DirectAU \cite{DirectAU_wang_2022_KDD} not only requires substantially longer training time, but also converges more slowly on both datasets. In contrast, DIAURec significantly reduces the overall training cost while jointly optimizing collaborative filtering representations and textual semantic representations, and even exhibits faster convergence on some datasets. Moreover, although DIAURec and IRLLRec \cite{2025IRLLRec} have comparable training time, IRLLRec typically relies on multiple calls with more complex prompts to construct language representations, which may introduce additional non-negligible overhead in practical deployments. Overall, DIAURec achieves a better efficiency trade-off while improving recommendation performance, making it more suitable for real-world recommendation scenarios.

\subsection{In-depth Analysis of DIAURec}
\subsubsection{\textbf{Ablation Analysis (RQ2)} }
\label{sec:ablation}To investigate the effect of each module, we construct the following variants: 

\begin{itemize}[leftmargin=*,label=--]
    \item \textbf{DIAURec\textsubscript{w/o FM}:} Remove the Fine-grained Matching approach and use the coarse-grained matching for optimization (Eq.~\ref{eq:fine_match}).  
    \item \textbf{DIAURec\textsubscript{w/o CM}:} Remove the Coarse-grained Matching approach and use the fine-grained matching for optimization (Eq.~\ref{eq:coarse_match}).  
    \item \textbf{DIAURec\textsubscript{w/o BothM}:} Remove the Both Matching approaches, include the fine- and coarse-grained matching.  
    \item \textbf{DIAURec\textsubscript{w/o DIIR}:} Remove the Dual-Intent reconstruction and the corresponding Intra-space Regularization (Eqs.~\ref{eq:reconstruction_gcn} and \ref{eq:intra_reg}).  
    \item \textbf{DIAURec\textsubscript{w/o IR}:} Remove the Interaction Regularization and consider only the intra-space regularization  (Eq.~\ref{eq:inter_reg}).  
    \item \textbf{DIAURec\textsubscript{w/o BothR}:} Remove the Both Regularization technologies, include the intra-space and interaction regularization.
\end{itemize}

\begin{table}
    \centering
        \footnotesize
    \caption{Ablation experiments of DIAURec \textit{w.r.t.} Recall@20 and NDCG@20 on the Amazon-book, Yelp, and Steam datasets.}
    \label{tab:ablation}
    \begin{tabular}{l|p{0.69cm}p{0.69cm}|p{0.69cm}p{0.69cm}|p{0.69cm}p{0.69cm}}
    \specialrule{0.75pt}{0pt}{0pt}
    \multirow{2}{*}{\textbf{Variants}} & \multicolumn{2}{c|}{\textbf{Amazon-book}} & \multicolumn{2}{c|}{\textbf{Yelp}} & \multicolumn{2}{c}{\textbf{Steam}} \\
    \cline{2-7} 
    & R@20 
    & N@20 
    & R@20 
    & N@20  
    & R@20 
    & N@20 \\
    \hline
    \hline
    w/o FM (Eq.\ref{eq:fine_match})
    &0.1619 &0.1018		
    &0.1311	&0.0843
    &0.1497	&0.0957  \\
    w/o CM (Eq.\ref{eq:coarse_match})
    &0.1654 &0.1044		
    &0.1298	&0.0846
    &0.1487	&0.0948  \\
    w/o BothM
    &0.1606 &0.1013		
    &0.1289	&0.0825
    &0.1454	&0.0929  \\
    \hline
    w/o DIIR (Eqs.\ref{eq:reconstruction_gcn}, \ref{eq:intra_reg})
    &0.1662 &0.1032		
    &0.1215	&0.0769
    &0.1341	&0.0835  \\
    w/o IR (Eq.\ref{eq:inter_reg})
    &0.1679 &0.1679		
    &0.1304	&0.0822
    &0.1516	&0.0961  \\
    w/o BothR
    &0.1615 &0.0999
    &0.1048 &0.0660
    &0.1340 &0.0834 \\
    \hline\hline
    \textbf{DIAURec}
    &\textbf{0.1701} & \textbf{0.1063}	
    &\textbf{0.1332} &\textbf{0.0847}
    &\textbf{0.1526} &\textbf{0.0971} \\
    \specialrule{0.75pt}{0pt}{0pt}
    \end{tabular}
\end{table}

We conducted a systematic ablation study on the core modules of DIAURec in Table \ref{tab:ablation}. 
The results clearly show that removing any module leads to performance degradation. 
For representation matching, eliminating either the fine-grained or coarse-grained component results in noticeable performance declines, while removing both simultaneously leads to the most severe degradation. This observation suggests that these two components capture complementary information at different granularities, both of which are essential for effective representation matching. 
Moreover, the results indicate that relying solely on alignment and uniformity is insufficient to fully bridge the discrepancies across heterogeneous representation spaces, highlighting the necessity of more comprehensive matching strategies.

Additionally, representation regularization proves to be indispensable. Specifically, removing all regularization techniques results in even larger performance drops, further demonstrating its critical role in mitigating representation collapse and stabilizing the training process. 
More importantly, these findings collectively suggest that effective representation learning should not only focus on enhancing the expressiveness of learned representations for better interpretability, but also place significant emphasis on optimization-related aspects, including the design of learning objectives, representation matching strategies, and regularization mechanisms, all of which are systematically integrated in DIAURec. 

\subsubsection{\textbf{Impact of Optimization Objectives (RQ3)}}

To validate the effectiveness of different optimization objectives, we trained DIAURec with alternative primary losses, as illustrated in Table \ref{tab:obj}. 
In $\text{DIAURec}_b$, we replaced the AU losses with the widely used Bayesian Personalized Ranking (BPR) loss and found that the performance dropped significantly. This degradation is caused not only by the false negatives introduced through random negative sampling in BPR, but also by its limited ability to guide effective representation optimization. 
In addition, we further analyze the limitations of uniformity optimization. To this end, we construct two variants: $\text{DIAURec}_{ui}$ imposes uniformity on both users and items, while $\text{DIAURec}_i$ applies it only to items. Experimental results demonstrate that item uniformity weakens the model’s ability to capture relations among similar items, ultimately leading to performance degradation. These results indicate that although uniformity is beneficial, over optimization on items can be counterproductive.

\begin{table}
    \centering
        \footnotesize
    \caption{Different optimization objectives of DIAURec \textit{w.r.t.} Recall@20 and NDCG@20 on the Amazon-book, Yelp, and Steam datasets.}
    \label{tab:obj}
    \begin{tabular}{l|p{0.69cm}p{0.69cm}|p{0.69cm}p{0.69cm}|p{0.69cm}p{0.69cm}}
    \specialrule{0.75pt}{0pt}{0pt}
    \multirow{2}{*}{\textbf{Method}} & \multicolumn{2}{c|}{\textbf{Amazon-book}} & \multicolumn{2}{c|}{\textbf{Yelp}} & \multicolumn{2}{c}{\textbf{Steam}} \\
    \cline{2-7} 
    & R@20 
    & N@20 
    & R@20 
    & N@20  
    & R@20 
    & N@20 \\
    \hline
    \hline
    \textbf{DIAURec}
    &\textbf{0.1701} & \textbf{0.1063}	
    &\textbf{0.1332} &\textbf{0.0847}
    &\textbf{0.1526} &\textbf{0.0971} \\
    $\text{DIAURec}_{b}$
    &0.1565 &0.0961		
    &0.1181	&0.0749 
    &0.1299	&0.0822\\
    $\text{DIAURec}_{u,i}$ 
    &0.1672 &0.1060		
    &0.1318	&0.0843
    &0.1441	&0.0915 \\
    $\text{DIAURec}_{i}$ 
    &0.1657 &0.1044		
    &0.1311	&0.0839
    &0.1422	&0.0897 \\
    \hline
    \specialrule{0.75pt}{0pt}{0pt}
    \end{tabular}
\end{table}

\subsubsection{\textbf{Impact of GCN Encoder (RQ4)}}
To examine the effectiveness of the GCN encoder in $\text{DIAURec}$, we conducted experiments with and without the encoder, as shown in Table \ref{tab:gcn}. 
Consistent with prior work, using only a base encoder limits representation optimization due to the absence of neighborhood information. In contrast, incorporating the GCN encoder leads to substantial performance improvements. 
Distinct from existing methods, we observe that $\text{DIAURec}$ reaches its optimal performance at 2 layer, after which additional layers result in declining performance. 
This drop can be explained by the over-smoothing \cite{wu_SCCF_KDD_2024} effect induced by high-order information. 
We argue that optimizing one-order neighborhood information is sufficient to ensure representation quality, whereas excessive high-order information leads to homogenization and ultimately undermines recommendation performance. 

\begin{table}
    \centering
        \footnotesize
    \caption{Different GCN layers of DIAURec \textit{w.r.t.} Recall@20 and NDCG@20 on the Amazon-book, Yelp, and Steam datasets.}
    \label{tab:gcn}
    \begin{tabular}{l|p{0.75cm}p{0.8cm}|p{0.75cm}p{0.8cm}|p{0.75cm}p{0.8cm}}
    \specialrule{0.75pt}{0pt}{0pt}
    \multirow{2}{*}{\textbf{Layer}} & \multicolumn{2}{c|}{\textbf{Amazon-book}} & \multicolumn{2}{c|}{\textbf{Yelp}} & \multicolumn{2}{c}{\textbf{Steam}} \\
    \cline{2-7} 
    & R@20 
    & N@20 
    & R@20 
    & N@20  
    & R@20 
    & N@20 \\
    \hline
    \hline
    $L=0$
    &0.1539 &0.0965		
    &0.0985	&0.0610 
    &0.0948	&0.0572\\
    \hline
    $L=1$
    &0.1669 &0.1044		
    &0.1312	&0.0835  
    &0.1498	&0.0954\\
    $L=2$
    &$\textbf{0.1701} \uparrow$ &$\textbf{0.1063}\uparrow$	
    &$\textbf{0.1332} \uparrow$ &$\textbf{0.0847}\uparrow$
    &$\textbf{0.1526} \uparrow$ &$\textbf{0.0971}\uparrow$ \\
    $L=3$
    &$0.1669 \, \downarrow$&$0.1040  \, \downarrow$	
    &$0.1297 \, \downarrow$&$0.0828  \, \downarrow$
    &$0.1484 \, \downarrow$&$0.0941  \, \downarrow$ \\
    $L=4$
    &0.1652 &0.1030		
    &0.1272 &0.0812
    &0.1463 &0.0925 \\
    \hline
    \specialrule{0.75pt}{0pt}{0pt}
    \end{tabular}
\end{table}

\subsubsection{\textbf{Impact of Hyperparameters (RQ5)}} 
\begin{figure}[!t]
  \centering

  \begin{subfigure}{0.49\linewidth}
    \centering
    \includegraphics[width=\linewidth]{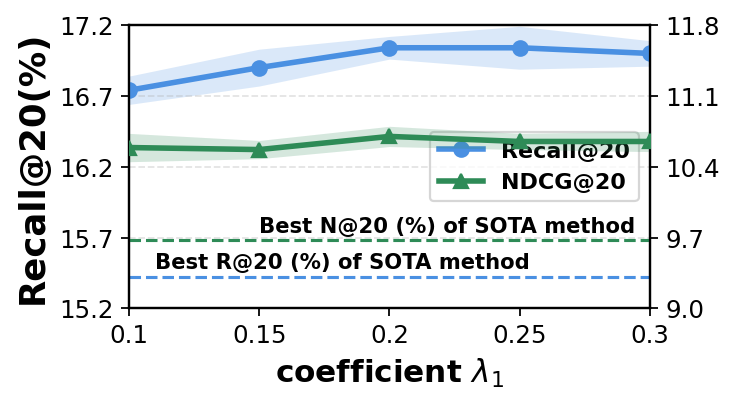}
  \end{subfigure}
  \begin{subfigure}{0.49\linewidth}
    \centering
    \includegraphics[width=\linewidth]{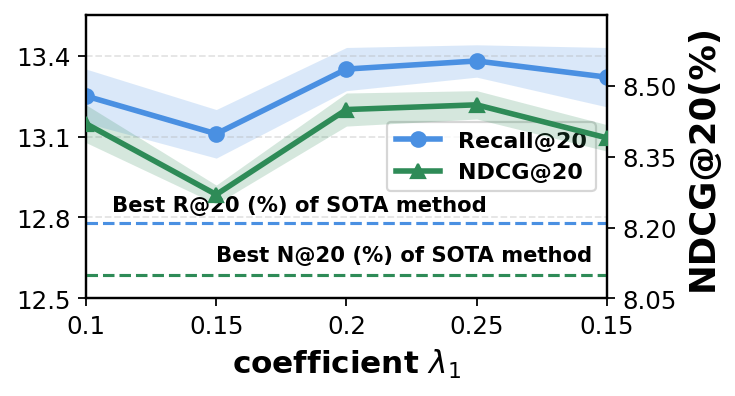}
  \end{subfigure}
  
  \begin{subfigure}{0.49\linewidth}
    \centering
    \includegraphics[width=\linewidth]{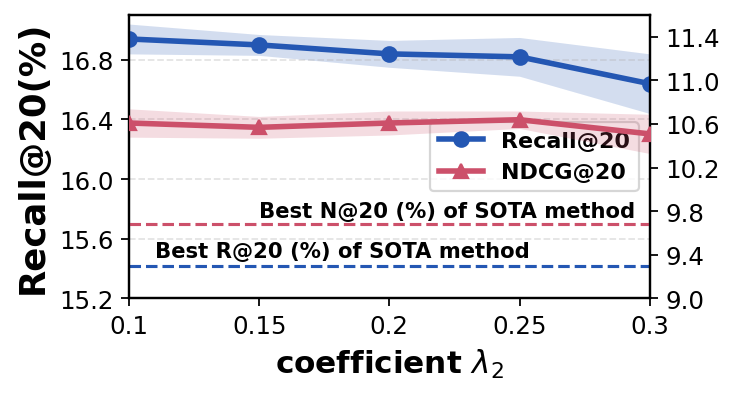}
    \caption{Amazon-book}
  \end{subfigure}
  \begin{subfigure}{0.49\linewidth}
    \centering
    \includegraphics[width=\linewidth]{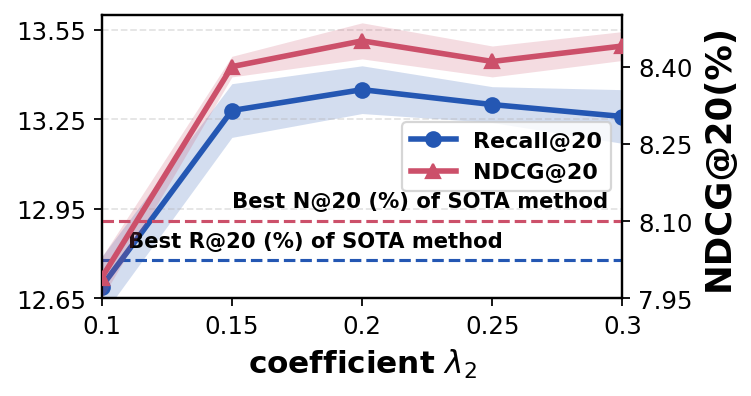}
    \caption{Yelp}
  \end{subfigure}

\caption{Hyperparameter sensitivities \textit{w.r.t.} Recall@20 and NDCG@20 for the matching weight $\lambda_1$ and regularization weight $\lambda_2$ on Amazon-book and Yelp datasets.}
  \label{fig:hyper}
\end{figure}

To optimize representation, in addition to the desired properties of alignment and uniformity, DIAURec incorporates representation matching approaches and regularization techniques. Although the ablation study in Section \ref{sec:ablation} verifies their effectiveness, these designs also introduce two hyperparameters. Fig. \ref{fig:hyper} reports the results under different settings. 
Overall, the performance of $\lambda_1$ and $\lambda_2$ remains relatively stable and consistently surpasses existing state-of-the-art methods, indicating that all modules steadily contribute to representation learning. 
However, on the Yelp dataset, when the weight $\lambda_2$ decreases to 0.1, model performance drops sharply due to representation space collapse. This result demonstrates that regularization plays a crucial role in preventing representation degradation during complex modeling and matching. 

\section{Related Work}

\noindent\textbf{Representation Modeling for Recommendation.} 
Effective representation modeling is essential for recommender systems \cite{wu_TOIS_RL_2019,RLMRec_2024_WWW,lu2026cluesgenerationlanguageguidedconditional}, since the representation's quality directly determines the accuracy of personalized ranking. 
Graph-based models, such as NGCF \cite{wang_SIGIR_NGCF_2019}, LightGCN \cite{he_LightGCN_SIGIR_2020}, and GMCF \cite{su_GMCF_SIGIR_2021}, explicitly leverage graph neural networks aggregate neighborhood information to refine representations. 
Beyond structural modeling, recent studies \cite{chen_WWW_IntentModeling_2022,DCCF_2023_SIGIR} emphasize intent modeling, which aims to capture the latent intentions behind user behaviors. 
Specifically, intent-based methods disentangle hidden user interests from observed actions, leading to more fine-grained and dynamic representations. 
Representative approaches include DCCF \cite{DCCF_2023_SIGIR}, BIGCF \cite{zhang_BIGCF_SIGIR_2024}, and IRLLRec \cite{2025IRLLRec}, which have verified the effectiveness of enriching collaborative signals with behavioral semantics. 
Additionally, large language models (LLM) \cite{bao_RS_TALLRec_2023,wu_LLMsurvey_2024} have also been leveraged to construct semantically enriched representations from textual attributes, reviews, or profiles. For instance, KAR \cite{KAR_2024_RS} integrates knowledge-aware LLM embeddings with collaborative signals, RLMRec \cite{RLMRec_2024_WWW} employs LLM to capture user intent from reviews, and AlphaRec \cite{AlphaRec_2025_ICLR} explores adaptive fusion of semantic and collaborative representations. 
While these methods effectively supplement collaborative signals with language representations, they often overlook how to collaboratively optimize representations from different modalities. This gap motivates our research on representation learning for recommendation.

\noindent\textbf{Representation Optimization for Recommendation.} 
Beyond effective representation modeling, the key challenge lies in how to appropriately exploit and refine these representations to maximize their utility. 
Self-supervised learning (SSL) \cite{gui_TPAMI_survey_2024,wu_SIGIR_SGL_2021} has recently emerged as a powerful paradigm to indirectly enhance representation quality. 
In recommendation, SSL has been explored with diverse augmentation strategies, leading to representative methods such as SGL \cite{wu_SIGIR_SGL_2021}, SimGCL \cite{SimGCL_yu_2022_SIGIR}, and LightGCL \cite{xu_LightGCL_ICLR_2023}. 
These studies show that SSL can improve robustness and alleviate data sparsity, thereby refining collaborative representation. 
Beyond indirect enhancement, recent work turns to direct optimization of representation space, particularly alignment and uniformity (AU) \cite{wang_ICML_understand_2020}. 
Alignment encourages positive pairs to be close, while uniformity ensures that the representation distribution remains well spread on the hypersphere. 
In recommender systems, DirectAU \cite{DirectAU_wang_2022_KDD}, CARec \cite{CARec_CIKM_2024}, and SIURec \cite{SIURec} exemplify this line of research, showing that AU-based methods can improve representation quality. 
However, these studies do not perform representation optimization from the perspective of language and intent modeling, and blindly aligning different representation spaces inevitably introduces noise. The proposed DIAURec stands in contrast to the design philosophy of prior works. This study introduce dual-intent modeling, reconstructing representations through the construction of prototype and distribution intents. In addition, coarse- and fine-grained matching approaches are introduced to achieve effective representation optimization. 

\section{Conclusion} 
In this paper, we proposed \textbf{DIAURec}, a dual-intent space representation learning framework that reconstructed representations by jointly modeling prototype intent and distribution intent, thereby disentangling users’ implicit preferences. 
To enhance representational consistency, we introduced alignment and uniformity objectives, combined with coarse- and fine-grained matching approaches as well as intra-space and interaction regularization. These designs collectively improved representation quality and effectively prevented collapse. Experiments on three highly sparse datasets showed that DIAURec consistently outperformed state-of-the-art baselines, fully validating its effectiveness and superiority.

In future work, we will plan to extend DIAURec to dynamic recommendation scenarios, such as sequential settings where user intents evolve over time. We will also explore incorporating richer side information ($e.g.,$ textual and multimodal features) to further enhance representation expressiveness. In addition, improving the scalability and adaptivity of the framework for large-scale and sparse environments remains an important direction.




\bibliographystyle{ACM-Reference-Format}
\bibliography{sample-base}


\end{document}